\documentclass[aps,twocolumn,prd,showpacs,nofootinbib]{revtex4}
\usepackage{amsmath}
\usepackage{graphicx}
\usepackage{dcolumn}
\usepackage{bm}
\usepackage{amssymb}
\usepackage{latexsym}

\def\be{\begin{equation}}
\def\ee{\end{equation}}
\def\ba{\begin{eqnarray}}
\def\ea{\end{eqnarray}}

\begin{document}

\title{Island Cosmology in the Landscape }

\author{Yun-Song Piao}
%$^{a,b}$}
%\affiliation{${}^a$  \rm CCAST(World Lab.), P.O.Box 8730, Beijing
%100080, China}

\affiliation{College of Physical Sciences, Graduate School of
Chinese Academy of Sciences, Beijing 100049, China}

%\email{yspiao@itp.ac.cn}
%\affiliation{${}^a$ Dept. of Physics, Graduate School of Chinese
%Academy of Sciences, YuQuan Road 19A, Beijing 100049, China}
%\affiliation{${}^a$Institute of High Energy Physics, Chinese
%Academy of Science, P.O. Box 918-4, Beijing 100039, P. R. China}
%\affiliation{${}^b$Interdisciplinary Center of Theoretical
%Studies, Chinese Academy of Sciences, P.O. Box 2735, Beijing
%100080, China}

\begin{abstract}

In the eternally inflationary background driven by the metastable
vacua of the landscape, it is possible that some local quantum
fluctuations with the null energy condition violation can be large
enough to stride over the barriers among different vacua, so that
create some islands full of radiation in new vacua, and then these
emergently thermalized islands will enter into the evolution of
standard big bang cosmology. In this paper, we calculate the
spectrum of curvature perturbation generated during the emergence
of island.
%, and show its spectral index and
%amplitude
We find that generally the spectrum obtained is nearly scale
invariant, which can be well related to that of slow roll
inflation by a simple duality. This in some sense suggests a
degeneracy between their scalar spectra. In addition, we also
simply estimate the non-Gaussianity of perturbation, which is
naturally large, yet, can lie in the observational bound well. The
results shown here indicate that the island emergently thermalized
in the landscape can be consistent with our observable universe.

\end{abstract}


\maketitle

\section{introduction}

%The spacetime background will be eternal inflating \cite{V1983,
%L1986}, when the cosmological dynamics is controlled by a
%landscape with large number of metastable minima. It is generally
In the eternally inflationary background driven by the metastable
vacua of the landscape \cite{V1983, L1986}, generally the bubbles
with new vacua will nucleate in original vacua \cite{S1983, GW},
which are mediated by the CDL instanton \cite{CDL}. However, these
bubbles are either empty or dominated by the new vacua. Thus in
order to obtain an observable universe, a slow roll inflation and
subsequent reheating has to be required inside the bubble.
%Thus it will be interesting to check whether there are other
%possibilities to lead to an universe like ours.
%There may be many quantum fluctuations with various spatial and
%temporal scales in each vacua of the landscape, which may violate
%the null energy condition \cite{W}, see also Ref. \cite{W0612} for
%discussions.
However, recently, it has been argued that some local quantum
fluctuations with the null energy condition violation might be
large enough to stride over the barrier among different vacua in
the landscape, so that straightly create many thermalized regions
in new vacua, some of which may correspond to our observable
universe \cite{Piao0706b}. The ``thermalized" here means that the
resulting state is a thermal state full of radiation and matter,
in which all components are assumed to be in thermal equilibrium.
Thus this part of region after the thermalization is quite similar
to that after the reheating following a slow roll inflation, and
so can be followed by a standard FRW evolution.

%These thermalized regions can be distinguished from the bubbles
%nucleated by the CDL instanton. The bubbles are either empty or
%dominated by the vacuum energy of new vacua, while here the
%thermalized regions are those straightly filled with radiation and
%matter.
These thermalized regions are referred as the ``islands'' in Ref.
\cite{Piao0706b}, which in some sense inherits but is actually
slightly different from the original idea of Ref. \cite{DV} and
Ref. \cite{Piao0506}, since here these islands are suggested to
origin from large fluctuations in parent vacuum but emerges in a
different or baby vacuum. It is this difference that makes the
emergence of the island here inevitably be related to the
tunneling in the landscape, especially the HM tunneling \cite{HM}.
We may phenomenally illustrate the island universe by applying the
HM instanton. The HM instanton corresponds a fluctuation which
makes the field jump to the top of the potential barrier, and then
the field will rapidly roll down along the another side of barrier
to the new vacuum and so it can be expected that the same
reheating as that after slow roll inflation will occur, see Fig.1.
In this sense, the emerging probability of the island in new vacua
can be approximately given by that of the HM instanton between
different vacua. There have been lots of studies on the HM
instanton, e.g. \cite{L1983, S1984} and recent \cite{L0611}. While
in Refs. \cite{DV, Piao0506}, it is not quite clear how to
calculate the emerging probability of an island. The island
universe should be distinguished from other models based on the
upward fluctuations, for examples, the study of Ref. \cite{DKS},
in which the state after the fluctuation is assumed to be an
observable universe with many structures, and also the recycling
universe proposed in Ref. \cite{GV}, see also Ref. \cite{LW1987}
and \cite{JS}, in which the state after the fluctuation is another
inflation state, see Ref. \cite{A07} for more discussions on
upward fluctuations .

\begin{figure}[t]
\begin{center}
\includegraphics[width=8cm]{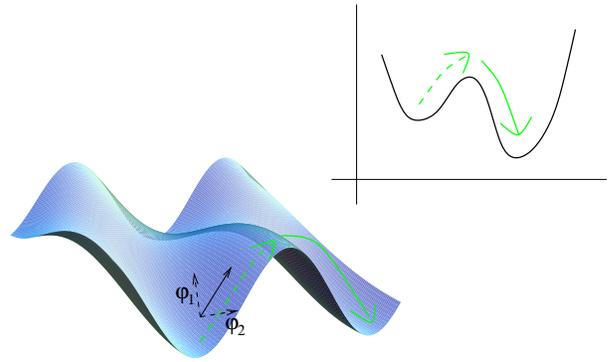}
\caption{The left down panel is the figure of a landscape with two
scalar fields, in which an upward fluctuation will inevitably
involve the fluctuations of both fields. To phenomenally simulate
this kind of fluctuation, we need to introduce two fields with the
reverse sign in their dynamical terms here. The right up panel is
its section, which is used to illustrate the island universe model
depicted here. The dashed line denotes an upward jump to the top
of the potential barrier, which can be realized by the HM
instanton or others, and then the field will rapidly roll down
along the another side of barrier to new vacuum, see the solid
line, and the same reheating as that after slow roll inflation
will be expected to occur. }
\end{center}
\end{figure}

The island in the landscape can share same remarkable successes
with the slow roll inflation. The reason is that the inflation can
be generally regarded as an accelerated or superaccelerated stage,
and so can be defined as an epoch when the comoving Hubble length
decreases, which actually occurs equally during an expansion with
the null energy condition violation.
%Then, after being located in top, the field will roll down rapidly
%towards new vacuum along the abrupt potential in another side, and
%then the reheating occurs, which corresponds to a thermalization
%making the island produced enter into the usual FRW evolution.
The emergence of the island leads to the decrease of the comoving
Hubble length, which makes the perturbations initially deep inside
the horizon be able to leave the horizon. The island universe
after the thermalization will enter into the usual FRW evolution.
During the period dominated by the radiation or matter the
comoving Hubble length is increasing, thus the perturbations on
super horizon scale will reenter into the horizon, and become
responsible for the structure formation of the observable
universe. Thus in this sense the emergence of the island naturally
leads to a solution to the horizon problem, and also the
generation of primordial perturbations. The total evolution can be
depicted in a causal patch diagram shown in Fig.2, see Refs.
\cite{Piao0506, Piao0512} for details.

In Ref. \cite{Piao0506}, the author has firstly calculated the
curvature perturbation of an island universe, in which the
background vacuum is taken as that with the observed value of
cosmological constant and the fluctuation is classically simulated
as that driven by a scalar field with a reverse sign in its
kinetic term, and it has been shown that the spectrum of metric
perturbation before the thermalization is dominated by an
increasing mode and is nearly scale invariant, which may induce
scale invariant curvature perturbation. However, whether the
resulting spectrum of curvature perturbation is scale invariant is
dependent of the physics at the epoch of thermalization. Thus in
this sense there is generally an uncertainty. The case here is
slightly similar to that for the bounce, see e.g. Refs. \cite{BF,
ABB}.

The emergence of the island in the landscape inevitably involves
the fluctuations of many fields, since the landscape can be
visualised as the space of a set of fields with a complicated and
rugged potential. Thus it will be more interesting to use more
scalar fields with the reverse sign in their kinetic terms to
phenomenally simulate the creation of island universe. In this
case there is not only the curvature perturbation but the entropy
perturbation. The curvature perturbation may be induced by the
entropy perturbation under certain condition, which may be
independent of the matching details at the thermalization surface.
This in some sense may relax the uncertainty of spectrum of
curvature perturbation leaded by the loophole of metric
perturbation propagating through the thermalization surface.

\begin{figure}[t]
\begin{center}
\includegraphics[width=7cm]{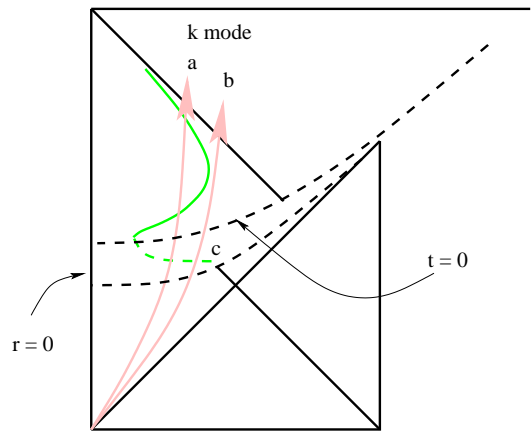}
\caption{The causal patch diagram of the creation and evolution of
island universe produced in a new vacuum distinguished from parent
vacuum, in which the green line and red lines denote the evolution
of comoving Hubble scale and primordial perturbation modes
respectively. The region between black dashed lines denotes the
fluctuation with null energy condition violation. The dashed line
of $t =0$ is the thermalization surface, after which the
fluctuation is over, the local island universe produced in new
vacuum starts its FRW evolution, and with the lapsing of time
finally will be dominated by new vacuum. When the energy density
of parent vacuum is high, there may be not enough efolding number
required by observable cosmology, which corresponds to the case
that the phase transition point `c' in figure is below the `a'
mode, while if the energy energy of parent vacuum is low enough,
there are always some perturbations never reentering and remaining
outside the Hubble scale after they leave from it during the
fluctuation, see `b' mode in figure, which means that in this case
a part of island is permanently inaccessible to any given observer
in island.}
\end{center}
\end{figure}

In this paper, we will present a detailed calculation for the
spectrum of curvature perturbation of the island universe emerged
in the landscape, which will be given in a more general case
making the predictions of our model more flexible. We find that
the results obtained can be well related to those of inflation by
a simple duality. In addition, we also simply estimate the
non-Gaussianity
of this curvature perturbation. % and compare it with the
%observation.
The final is our discussion.

%Then this thermalized region will evolve with standard cosmology.

\section{island in the landscape}

In this section, we briefly show some model independent characters
of the island universe in the landscape. Though the spawning of
island in the landscape is actually a quantum process, it can be
regarded phenomenally or semiclassically as an evolution with the
null energy condition violation to study, as was done in Ref.
\cite{Piao0506}, see also earlier work in Ref. \cite{PZ}. When the
island emerges, the change of local background may be depicted by
the drastic evolution of local Hubble parameter `$h$', where the
``local" means that the quantities, such as the scale factor `$a$'
and `$h$', only character the values in null energy condition
violating region. Then we introduce the parameter `$\epsilon$'
defined as $-\,{\dot h}/ h^2$, which
%can be regarded as $\epsilon
%\simeq {1\over h \Delta t}{\Delta h\over h}$, and thus actually
describes the change of $h$ in unit of Hubble time. For the null
energy condition violating fluctuations, ${\dot h}>0$, thus
$\epsilon <0$ can be deduced. The simplest selection for the scale
factor $a(t)$ is given by \be a(t) \sim e^{\int{n\over (-t)}dt}
%\,\,\,\,\,\,h={n\over (-t)}
\label{h}\ee where $t$ is negative and runs up to $0$, and $n$ is
a positive parameter dependent of time. We can obtain $\epsilon=
-{1\over n}+{dn\over (nh)dt}$,
%where $h$ given by Eq.(\ref{h}) has
%been used,
which will be used in the calculations of primordial
perturbations. In addition, we also assume $ |{1\over n}|\gg
|{dn\over (nh)dt}|$ in this paper, which implies $|{dn\over
h(dt)}|\ll 1$, otherwise the perturbation spectrum will be not
nearly scale invariant. This means that the change of $n$ in unite
of Hubble time must be quite small. The more rapid the fluctuation
is, in principle the stronger it can be \cite{DV}, which in some
sense is also a reflection of the uncertainty relation between the
energy and time in quantum dynamics. Thus to make the fluctuate be
strong enough to create the islands of our observable universe,
%we
%should take the time scale of the NEC violation \be \int dt =
%\int_i^e {dh \over |\epsilon | h^2} \simeq {1\over |\epsilon |
%h_i} \label{indt}\ee be vanishingly small, where the subscript `i'
%and `e' denote the initial and end value of the NEC violating
%fluctuation respectively, and thus $h_i$ is determined by the
%energy scale of original vacuum. To make $\int dt \rightarrow 0$,
%we need $h_i$ or $|\epsilon |$ is quite large. The larger $h_i$ is
%the more violent the fluctuation is anticipative. However, as will
%be showed here, $h_i$ is required to be so small as to solve the
%horizon problem of standard cosmology. Thus Eq.(\ref{indt})
%suggests
$|\epsilon |\gg 1$ is required \cite{Piao0706b, Piao0506}. Thus
though during the fluctuation the change of $h$ is drastic, the
expansion of the scale factor is extremely slow, since we have
$a\sim (-t)^{1\over |\epsilon|}$ from Eq.(\ref{h}) for $n$
approximately constant.

%\begin{figure}[t]
%\begin{center}
%\includegraphics[width=7cm]{islandpen.eps}
%\caption{The causal patch diagram of the creation and evolution of
%island universe produced in a different vacuum, in which the green
%line and red lines denote the evolution of Hubble scale and
%primordial perturbation modes respectively. The region between
%black dashed lines denotes the NEC violating fluctuation. The
%dashed line of $t =0$ is the thermalization surface, after which
%the NEC violating fluctuation is over, the local island universe
%produced in new vacuum starts its FRW evolution, and with the
%lapsing of time finally will be dominated by new vacuum. }
%\end{center}
%\end{figure}

%In the island, the horizon problem of standard cosmology can be
%solved and the generation of primordial perturbation can be
%implemented are as follows.

%The perturbations leaving the horizon
%during the NEC violating fluctuation might reenter the horizon
%during radiation and matter domination after the end of
%fluctuation, which may be responsible for the structure formation
%of our observable universe, see Fig.2 for an illustration. For our
%purpose it may be convenient to define \be {\cal N} \equiv
The efolding number of mode with some scale $\sim 1/k$ leaving the
horizon before the thermalization can be defined as \be {\cal
N}=\ln{\left({a_eh_e\over ah}\right)}, \label{caln}\ee where the
subscript `e' denotes the quantity evaluated at the time of the
thermalization, and thus $k_e=a_eh_e$ is the last mode to be
generated. When taking $ah=a_0h_0$, where the subscript `0'
denotes the present time, we generally have ${\cal N}\sim 50$,
which is required by observable cosmology.
%, see Fig.2 for an illustration with the cause patch
%diagram.
%The evolution of perturbations in the island universe
%model may be depicted in a causal patch diagram, see Fig.2. We can
%see that if the energy value of original vacuum is high, there may
%be not enough efolding number required by observable cosmology,
%which corresponds to the case that the phase transition point `c'
%in Fig.2 is below the `a' mode, while if the energy value of
%original vacuum is low enough, there are always some perturbations
%never reentering and remaining outside the Hubble scale after
%their leaving from it during the null energy condition violating
%fluctuation, see `b' mode in Fig.2, which means that in this case
%a part of island is permanently inaccessible to any given observer
%in the island.
By using Eq.(\ref{caln}), and also note that the scale factor $a$
is nearly unchanged here, we can obtain \be {\cal N}\cong
\ln\left({h_e\over h_i}\right)\cong \ln\left({T_e\over
\Lambda_i}\right)^2, \nonumber\ee where $\Lambda_i\simeq
h_i^{1/2}$ is the energy scale of original or parent vacuum and
${T}_e\simeq h_e^{1/2}$ is the thermalized temperature after the
fluctuation, which is the same as the reheating temperature after
the inflation, and also the constant $G\equiv 1$ is set in this
paper. When taking ${\cal N}\simeq 50$ and ${T}_e \sim
10^{15}$Gev, we have $\Lambda_i\sim$ Tev. For a lower $\Lambda_i$,
${T}_e$ may be taken smaller. Thus it seems that the above
condition can be satisfied easily, which indicates that the
efolding number required to solve the horizon problem of standard
cosmology may be always obtained by selecting a low parent vacuum.
The reason is that the smaller the energy scale of parent vacuum
is, the larger its Hubble scale is, thus the efolding number, see
Fig.2 for a further illustration.
%This result also suggests that if our universe is actually
%such an island, then the observations made in our universe might
%have recorded some information on last vacuum. For example, for
%the efolding number ${\cal N}>50$, we may know the its energy
%scale should be low.
In principle, $T_e$ should be lower than that the monopoles
production needs, while higher than Tev. This can not only avoid
the monopole problem afflicting the standard cosmology, but helps
to provide a solution to the baryogenesis.

This can be also explained as follows. The scale of the null
energy condition violating region is generally required to be
larger than the Hubble scale of original vacuum \cite{DV}, see
also Refs. \cite{FG} and \cite{VT, BTV}. This is also assured by
the application of the HM instanton action, since in which the
region tunnelling to the top of the barrier corresponds to the
Hubble scale of the original vacuum, which have been understood by
using the stochastic approach to inflation \cite{AAS, GLM}. This
sets the initial value of local evolution of $a$, and since it is
nearly unchanged during the fluctuation, we can have $a_e\simeq
a_i\simeq 1/h_i$, which means that the smaller $h_i$ is, the
larger the scale of local thermalized region after the fluctuation
is, and thus the efolding number.

%\begin{figure}[t]
%\begin{center}
%\includegraphics[width=8cm]{islandpen.eps}
%\caption{The causal patch diagram of the creation and evolution of
%island universe produced in a different vacuum, in which the green
%line and red lines denote the evolution of comoving Hubble scale
%and primordial perturbation modes, respectively. The region
%between black dashed lines denotes the fluctuation with the null
%energy condition violation. The dashed line of $t =0$ is the
%thermalization surface, after which the fluctuation is over, the
%local island universe produced in new vacuum starts its FRW
%evolution, and with the lapsing of time finally will be dominated
%by new vacuum. }
%\end{center}
%\end{figure}

%The change of $h$ can be regarded as $h= {n\over -t}$, where
%generally $n$ may be slightly dependent of time. Thus \be\epsilon
%= -{1\over n}+{{\dot n}\over  nh}. \label{h}\ee  When $t$ is
%initially from $-\infty$ to $0-$, it corresponds to a slow
%expansion.

%The $\epsilon$ can be rewritten as $\epsilon \simeq {1\over h
%\Delta t}{\Delta h\over h}$, thus in some sense $\epsilon$
%actually describes the change of $h$ in unit of Hubble time and
%depicts the abrupt degree of background. From Eq.(\ref{h}), during
%the slow expansion, though the scale factor is hardly changed, the
%Hubble parameter will rapidly increase, which means an abrupt
%change of background
\section{Spectrum of Primordial Perturbation in Island}

\subsection{The calculations of curvature perturbation}

In this section, we will study the primordial perturbations of
island cosmology in the landscape, and, without loosing
generality, will discuss the case with the scalar fields
$\varphi_1$ and $\varphi_2$ with the reverse sign in their
dynamical terms, see Fig.1. In Ref. \cite{Piao0705}, the relevant
calculations have been done for constant $\epsilon$, however, here
we will relax the assumption of constant $\epsilon$ and allow it
change, which will make the results more flexible for matching to
the observations. We assume that both fields are uncoupled and
regard their potentials not around the top as \be V(\varphi_1)=
{n_1(3n+1)\over 8\pi } \exp{\left(-\int \sqrt{16\pi \over n_1}
d\varphi_1\right)},\label{vv1}\ee \be V(\varphi_2)=
{n_2(3n+1)\over 8\pi} \exp{\left(-\int\sqrt{16\pi \over n_2}
d\varphi_2\right)}, \label{vv2}\ee respectively, where the
determination of prefactors has been showed in Ref.
\cite{Piao0705}, and in general both $n_1$ and $n_2$ are positive
parameters dependent of the time in different forms. Note that in
Ref. \cite{Piao0705}, $n_1$ and $n_2$ were regarded as constants,
thus in Eqs.(\ref{vv1}) and (\ref{vv2}) the integrals in the
exponents can be integrated, which actually makes Eqs.(\ref{vv1})
and (\ref{vv2}) have more simple forms. Here in order to have a
detailed compare of spectrum with the observation we regard $n_1$
and $n_2$ be changed, thus the integrals in the exponents must be
reserved. However, to have a simple equation of entropy
perturbation, $n_1/n_2$ is constant is assumed here, which means
that their changes with the time are same, and also $n_1+n_2=n$,
where $n$ is given by Eq.(\ref{h}). Note that $n\ll 1$, thus both
$n_1, n_2\ll 1$, which suggests that the potential of both fields
are very steep in Eqs.(\ref{vv1}) and (\ref{vv2}). The fields
$\varphi_1$ and $\varphi_2$ during their evolution will climb up
along their potentials, which is determined by the property of
such fields, see e.g. Refs. \cite{ST, GPZ}. Thus in this sense
they can be suitable for simulating the emergence of island in the
landscape, since the emergence of island here actually corresponds
to an upward fluctuation along the potential in the landscape.

%Note also that it may be showed that this solution is an
%attractor, e.g. see Ref. \cite{GPCZ}.

We can decompose both fields into the field $\varphi$ along the
field trajectory, and the field $s$ orthogonal to the trajectory
by making a rotation in the field space \be d\varphi =
 {\sqrt{n_1}d\varphi_1+\sqrt{n_2}d\varphi_2\over
\sqrt{n}}, ds = {\sqrt{n_2}d\varphi_1-\sqrt{n_1}d\varphi_2\over
\sqrt{n}}, \nonumber\label{phis}\ee as has been done in Ref.
\cite{GWBM}. In this case, the total potential $V(\varphi,s)$,
which is the sum of Eqs.(\ref{vv1}) and (\ref{vv2}), can be
rewritten as ${\tilde V}(s) e^{-\int d\varphi\sqrt{16\pi \over n}}
$, where \ba {\tilde V}(s) & = & {n_1(3n+1)\over
8\pi} \exp{\left(-\int\sqrt{16n_2\pi \over n_1 n} ds\right)}\nonumber\\
& + & {n_2(3n+1)\over 8\pi} \exp{\left(\int \sqrt{16n_1\pi  \over
n_2 n} ds\right)}\label{vphi1}\ea is the potential of $s$ field,
whose effective mass is given by $\mu^2(s)= {\tilde
V}^{\prime\prime}(s)$. Thus we have \ba {\mu^2(s)\over h^2} &
\cong & {2+6n-{3\over \sqrt{16\pi n}}{dn\over d\varphi}\over
n^2}\nonumber\\ &\cong & {2- {6\over \epsilon} -{5\over
2}{{d\ln{|\epsilon |}}\over d{\cal N}}\over \left({1\over
\epsilon}\right)^{2}}, \label{mus}\ea where the high order terms
like $\left({dn\over d\varphi}\right)^2$ and ${d^2n\over
d\varphi^2}$ have been neglected, and in the second line the
higher order terms like $\left({d\ln{|\epsilon|}\over d{\cal
N}}\right)^2$ and ${d^2\ln{|\epsilon|}\over d{\cal N}^2}$ also
have been neglected. We educe the second line of Eq.(\ref{mus}) by
using
%noting that ${3\over \sqrt{16\pi n}}{dn\over d\varphi} =
%-{3\over 2}\sqrt{{|\epsilon| \over n}}\left({{d n}\over d{\cal
%N}}\right)$, where
the definition of $\epsilon$ and ${\cal N}$, and also
%noting that combining Eq.(\ref{h}) and the definition of
%$\epsilon$ can give $-{1\over n}\simeq \epsilon -{d\epsilon\over
%d{\cal N}}$,
noting that ${d\epsilon\over d{\cal N}}$ is far smaller than
$\epsilon$, since $ |{1\over n}|\gg |{dn\over (nh)dt}|$, as has
been assumed. Thus we see that Eq.(\ref{mus}) is not dependent of
$n_1$ and $n_2$, but only the background parameter $n$ or
$\epsilon$.

The perturbations will be decomposed into both parts after this
rotation is done, one is the curvature perturbation induced by the
fluctuation of $\varphi$ field,
%which actually
%dominates the evolution of background with the potential given by
%\be V(\varphi)\equiv {n(3n+1)\over 8\pi}
%\exp{\left(-\int\sqrt{16\pi \over n}d\varphi\right)},
%\label{vphi1}\ee
and the other is the entropy perturbation induced by the
fluctuation of $s$ field. In linear order, as long as the
background trajectory remains straight in field space, the entropy
perturbation must be decoupled from the curvature perturbation,
which actually can be seen in Ref. \cite{GWBM}. When the entropy
perturbation is decoupled from the curvature perturbation, the
calculation of curvature perturbation is the same as that in Ref.
\cite{Piao0506}, in which only when the increasing mode of metric
perturbation before the thermalization may be inherited by the
constant model of the curvature perturbation $\zeta$ after the
thermalization, the spectrum is scale invariant, or the spectrum
will be strong blue, whose amplitude is negligible on large scale.
In this case the entropy perturbation $\delta s$ may be calculated
as follows. In the momentum space, the equation of entropy
perturbation can be given by \be
v_k^{\prime\prime}+\left(k^2-{\beta(\eta)\over \eta^2}\right)v_k
=0, \label{uki}\ee where $\delta s_k\equiv v_k/a$ has been defined
and the prime denotes the derivative for the conformal time
$\eta$, and $\beta(\eta)$ is given by the sum of
${a^{\prime\prime}\over a}$ and $\mu^2a^2$, between which is not
subtraction sign as usual, since the fields used here have the
reverse sign in their dynamical terms, in which $\mu^2$ is
determined by Eq.(\ref{mus}). Note that for ${d\ln{|\epsilon
|}\over d{\cal N}}\ll 1$, $\beta$ is actually near constant for
all interesting modes $k$. Thus Eq.(\ref{uki}) is a Bessel
equation and its general solutions are the Hankel functions with
the order $v$ given by \ba v &=& \sqrt{\beta+{1\over
4}}\nonumber\\ &\cong & {3-{2\over
\epsilon}+{d\ln{|\epsilon|}\over d{\cal N}}\over 2}, \label{v}\ea
where ${1\over ah}\cong \eta\epsilon(1-{1\over
\epsilon}-{d\ln{|\epsilon|}\over d{\cal N}})$, which has been
given in Ref. \cite{LMTS}, has been used.

%&\simeq & {2-{3\over \epsilon}+{3\over 2}{d\ln{|\epsilon|}\over
%d{\cal N}}\over \eta^2},\label{beta}\ea

%where $\mu^2$ is given by Eq.(\ref{mus}), and
%This result is exactly same with that in Ref. \cite{Piao0705} when
%$\epsilon$ and thus $n$ is taken as the constant. in which we let
%$\beta$ be the denominator of Eq.(\ref{beta}).

%\ba f(\eta) &\equiv &
%{a^{\prime\prime}\over a}+\mu^2(s)a^2\nonumber\\
%&\simeq & {2-{3\over \epsilon}+{3\over 2}{d\ln{|\epsilon|}\over
%d{\cal N}}\over \eta^2},\label{beta}\ea

%where $\mu^2$ is given by Eq.(\ref{mus}), and ${1\over ah}\cong
%\eta\epsilon(1-{1\over \epsilon}-{d\ln{|\epsilon|}\over d{\cal
%N}})$, which has been give in Ref. \cite{LMTS}, has been used.
%This result is exactly same with that in Ref. \cite{Piao0705} when
%$\epsilon$ and thus $n$ is taken as the constant. The right side
%of the first line in Eq.(\ref{beta}) is the plus between two
%terms, but not minus as usual, which is actually the result using
%the fields with the reverse sign in their dynamical terms. Note
%that for ${d\ln{|\epsilon |}\over d{\cal N}}\ll 1$, $\beta$ is
%near constant for all interesting modes $k$, in which we let
%$\beta$ be the denominator of Eq.(\ref{beta}). Thus Eq.(\ref{uk})
%is a Bessel equation and its general solutions are the Hankel
%functions with the order $v$ given by \ba v &=&
%\sqrt{\beta+{1\over 4}}\nonumber\\ &\simeq & {3\over 2}-{1\over
%\epsilon}+{1\over 2}{d\ln{|\epsilon|}\over d{\cal N}}, \ea where
%Eq.(\ref{beta}) has been used.

In the regime $k\eta \rightarrow \infty $, all interesting modes
are very deep inside the horizon of the parent vacuum, thus
Eq.(\ref{uki}) can be reduced to the equation of a simple harmonic
oscillator, in which $v_k \sim e^{-ik\eta} /(2k)^{1/2}$.
%which in some sense suggests that the initial condition can be
%taken as that same with usual Minkowski vacuum.
In the superhorizon scale, i.e. $k\eta\rightarrow 0$, in which the
modes become unstable and increases, the expansion of Hankel
functions to the leading term of $k$ gives \be v_k\simeq {1\over
\sqrt{2k}}(-k\eta)^{{1\over 2}-v} ,\label{uk}\ee where the phase
factor has been neglected. The emergence of island goes with the
abrupt change of $h$, as has been mentioned in last section. Thus
it may be expected that the perturbation amplitude of $v_k$ will
continue to change after it leaves the horizon, up to the
thermalization epoch. This can also be explained as follows. To
make the analysis simplified, we assume $|\epsilon|$ is constant.
When $k\eta\rightarrow 0$, which corresponds to the super horizon
scale, from Eq.(\ref{uki}), we have $
v_k^{\prime\prime}-{\beta(\eta)\over \eta^2}v_k\simeq 0$. This
equation has one increasing solution and one decay solution. The
increasing solution is given by $ v_k\sim a^{|\epsilon|}$, see
Refs. \cite{Piao0608, Piao0705} for details. The scale factor $a$
is nearly unchanged, but since $|\epsilon|\gg 1$, the change of
$v_k$ has to be significant, thus generally one can not obtain
that the $|\delta s_k |=|v_k/a|\sim a^{|\epsilon|}$ is constant,
which actually occurs only for the slow roll inflation in which
approximately $|\epsilon|\simeq 0$.
%The details can also be seen in Ref. \cite{Piao0608}, in which the
%spectrum of a test scalar field not affecting the evolution of
%background was calculated, which in some sense corresponds to the
%case of $n_2=0$ here.
This result indicates that we should take the value of $v_k$ at
the time of thermalization to calculate the amplitude of
perturbations.
Thus the perturbation spectrum of entropy perturbation is \be %k^{3/2} |\delta s_k| \equiv
k^{3/2}{\cal P}^{1/2}_s =k^{3/2} \left|{v_k(\eta_e)\over a}\right|
%\left({n+1\over
%n}\right)^{2v-1}\left({h_e\over 2\pi}\right)^2\left({k\over k_e}\right)^{3-2v}\nonumber\\
%&\simeq &\left({1\over n}\right)^{2v-1}\cdot\left({h_e\over
%2\pi}\right)^2\cdot \left({k\over k_e}\right)^{3-2v}\nonumber\\
\sim k^{3/2-v}. \label{ps}\ee Thus the spectrum index is given by
\be n_s-1 \simeq {2\over \epsilon}-{d\ln{|\epsilon|}\over d{\cal
N}}, \label{ns}\ee where Eq.(\ref{v}) has been used, which means
that the spectrum of entropy perturbation generated during the
emergence of island is nearly scale invariant, since
$|\epsilon|\gg 1$ and ${d\ln{|\epsilon|}\over d{\cal N}}\ll 1$,
with a possible tilt determined by the evolution of background.
When $|\epsilon|$ is constant, Eq.(\ref{ns}) will be exactly back
to that in Ref. \cite{Piao0705}.

%during the slow expansion, but not dependent of other details.

The spectrum of entropy perturbation can be inherited by the
curvature perturbation,
%before the thermalization is able to
%inherit the generated, the entropy perturbation may be responsible
%well for the structure formation of observable universe.
which can be accomplished by noting that the entropy perturbation
sources the curvature perturbation by \be |{\dot \zeta}|\simeq
{2h{\dot\theta}\over {\dot\varphi}}\delta s \label{dotxi}\ee on
large scale \cite{GWBM}, where $\theta\equiv {\rm
arctg}\sqrt{n_2\over n_1}$ depicts the motion trajectory of both
fields in field space. When $\theta$ is a constant, it is a
straight line. In this case, ${\dot\theta}=0$, thus the entropy
perturbation will be not decoupled to the curvature perturbation,
which also assures the validity of Eq.(\ref{uki}), or there will
be some terms such as $\sim {\dot \theta}^2$ and $\sim {\dot
\theta}\Phi$. However, if there is a sharp change of field
trajectory, ${\dot\theta}$ must be not equal to $0$, in this case
$\zeta$ will inevitably obtain a corresponding change induced by
$\delta s$ by Eq.(\ref{dotxi}), as has been pointed out and
applied in ekpyrotic cosmology \cite{LMTS, BKO}, see also earlier
Refs. \cite{NR, DFB} and recent studies \cite{KW} on the ekpyrotic
collapse with multiple fields.

It may be expected that at the split second before the
thermalization $n_1/n_2$ will be not constant any more, since
around this epoch both fields will be in the top of their
potentials, and thus Eqs.(\ref{vv1}) and (\ref{vv2}) depicting the
upward fluctuation along the potential are generally not valid any
more. In this case, the entropy perturbation will be able to
source the curvature perturbation.
%which actually may be constructed by modifying the potential of
%fields around the . For example, around the end epoch, instead of
%being steep, the potential of one of fields will has a maximum or
%a plateau, which will lead to the rapid stopping of up climbing of
%corresponding field, while the up climbing of another field
%remains, note that here the motion of field is mainly managed by
%its potential, see e.g. Refs. \cite{ST, GPZ}.
We assume, for a brief analysis, that before the thermalization
the motion of $\varphi_2$ firstly rapidly stops while the other
field $\varphi_1$ remains and then will stop moving after several
split seconds. Following Ref. \cite{LMTS, BKO}, this corresponds
to a sharp change from initial value $\theta_{*}={\rm
arctg}\sqrt{n_2\over n_1}$ to $\theta\simeq 0$. It is this change
that leads $\zeta$ acquiring a jump induced by the entropy
perturbation and thus inherits the nearly scale invariant spectrum
of the entropy perturbation. In the rapid transition
approximation, one has obtained \be |\zeta | \simeq
{2h\Delta\theta\over {\dot\varphi}}\delta s
%\cdot\left({\rm arctg}\sqrt{n_2\over
%n_1}\right)
%\simeq {h_e\over {\dot\varphi}}\delta s
, \label{xi}\ee where $\Delta\theta \simeq\theta_{*}={\rm
arctg}\sqrt{n_2\over n_1}$. The amplitude of entropy perturbation
can be calculated at the time of thermalization and given by \be
{k^{3/2}\over \sqrt{2\pi^2}} \left|{v_k(\eta_e)\over a}\right|
%&\simeq & \left({n+1\over n}\right)^2\cdot\left({h_e\over
%2\pi}\right)^2 \nonumber\\
\simeq
%|\epsilon|\left({h_e\over 2\pi}\right)
%\left|1-{1\over\epsilon}-{d\ln{|\epsilon|}\over d{\cal
%N}}\right|\nonumber\\
%&\simeq &
|\epsilon|\left({h_e\over 2\pi}\right), \label{psa}\ee where
$|\epsilon|\gg 1$ and ${d\ln{|\epsilon|}\over {\cal N}}\ll 1$ have
been used.
%The calculations are similar to that done in
%Ref. \cite{Piao0608}. The prefactor $1/n$ is from the relation
%$1/\eta_e= (1+1/n) a_eh_e$, which corresponds the $g$ factor
%introduced and discussed in Ref. \cite{Piao0608}.
Note that $|h^2/{\dot\varphi}^2|= {4\pi\over |\epsilon|} $, thus
we have the amplitude of curvature perturbation \ba {\cal
P}_{(s\rightarrow \zeta)}& =& {k^3\over 2\pi^2}|\zeta^2_k|\cong
\left|{h\Delta\theta\over {\dot\varphi}}\right|^2\cdot {k^3\over
2\pi^2} \left|{v_k(\eta_e)\over a}\right|^2 \nonumber\\ &\simeq &
(2\Delta\theta)^2\cdot |\epsilon| {h_e^2\over \pi}. \label{pxi}\ea
which is approximately ${\cal P}_{(s\rightarrow \zeta)}\simeq
|\epsilon|h_e^2$.

\subsection{The duality of scalar spectra between inflation and
island}

We can note that Eqs.(\ref{ns}) and (\ref{pxi}) can be related to
those of the usual slow roll inflation by replacing $\epsilon$ as
$-{1\over \epsilon}$, which actually exactly gives the spectral
index and amplitude of slow roll inflation to the first order of
slow roll parameters. This replacement may be regarded as a
duality between their scalar spectra, which, in some sense, is a
reflection of duality between their background evolutions, i.e.
the nearly exponent expansion with $\epsilon\simeq 0$ and the slow
expansion with $\epsilon\ll -1$, see Fig.3. In Ref.
\cite{Piao0705}, we showed that this duality is valid for constant
$|\epsilon|$, here we find that it is still valid when
$|\epsilon|$ changes. This result extends again the studies on the
dualities of the primordial density perturbation in Refs.
\cite{KST, BST}, which discussed the cases of $\epsilon>0$, and
\cite{Piao, Lid}, in which the case of $\epsilon<0$ is included.

The duality between the scalar spectra of inflation cosmology and
island cosmology indicates that for a slow roll inflation model,
we can deduce $\epsilon({\cal N})$ by studying the details of
model to calculate its spectrum, however, by the dual relation
showed here we also always can write down a dual $\epsilon({\cal
N})$ for island universe model, both give same scalar perturbation
spectra. For instance, for large field inflation model, we have
$\epsilon\sim {1\over {\cal N}}$, then in term of duality
$\epsilon \rightarrow -{1\over \epsilon}$, we can take
$\epsilon\sim -{\cal N}$ for island universe, one will find that
in this case the scalar spectra of both models will be exactly
same. Thus in the level of scalar spectrum, the island universe
model is actually degenerated with the slow roll inflation model.

\begin{figure}[t]
\begin{center}
\includegraphics[width=7.5cm]{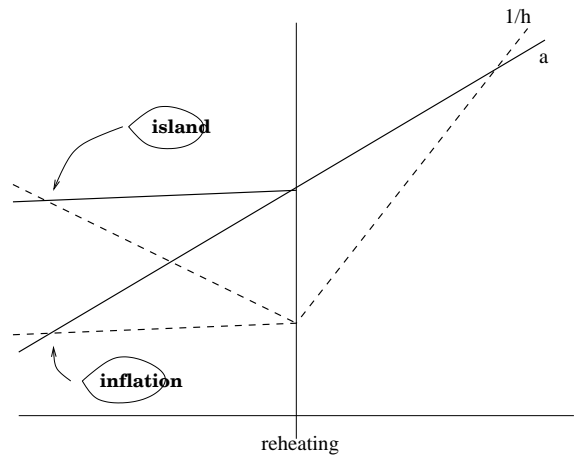}
\caption{The figure of the background evolutions of slow roll
inflation and island universe model. The black solid line is the
scale factor $a$, while the black dashed line is the horizon
radius $1/h$. For slow roll inflation, $a$ is rapidly increased
while $h$ is nearly unchanged. For island, $a$ is nearly unchanged
while $h$ is rapidly increased. The duality between their scalar
spectra is a reflection of that between their background
evolutions. }
\end{center}
\end{figure}

\subsection{The analysis of spectral tilt}

\begin{figure}[t]
\begin{center}
\includegraphics[width=8cm]{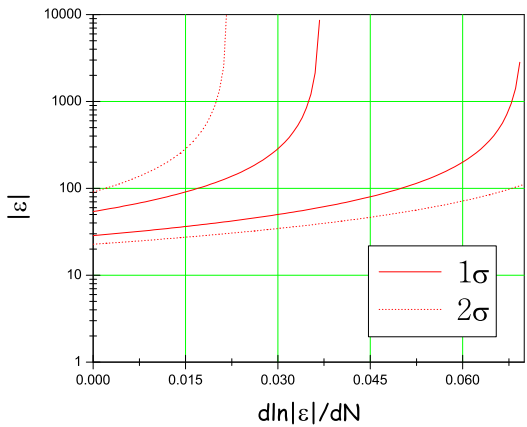}
\caption{ The region of $|\epsilon|$ with respect to
${d\ln{|\epsilon|}\over d{\cal N}}$. The $1\sigma$ and $2\sigma$
levels are given by WMAP$+$SDSS. We can see that the region of
parameter space consistent with the observations is quite large. }
\end{center}
\end{figure}

The tilt of spectral index given by Eq.(\ref{ns}) is interesting
for the observations. When $\epsilon$ is constant, the spectrum is
slightly red, since ${d\ln{|\epsilon|}\over d{\cal N}}=0$ and
$\epsilon \lesssim 0$, which may be matched to the observations
well. For instance, taking $|\epsilon|\simeq 50$, we have
$n_s\simeq 0.96$, which is well in the region favored by the
observation \cite{WMAP}. However, it may be more possible that
$\epsilon$ is not a constant. In this case, it may be expected
that the term ${d\ln{|\epsilon|}\over d{\cal N}}$ will
significantly affect the spectral index. When $|\epsilon |$ is
decreased with the decrease of ${\cal N}$, we have
${d\ln{|\epsilon|}\over d{\cal N}}>0$, which will make the
spectrum redder. For instance, taking $|\epsilon|\simeq {\cal N}$
at the epoch of ${\cal N}\simeq 50$, we have
${d\ln{|\epsilon|}\over d{\cal N}}={1\over {\cal N}}$ and thus
$n_s\simeq 1-{3\over {\cal N}}\simeq 0.94$. It, of course, is also
possible to have a mild red tilt by having a mild running of
$\epsilon$ with ${\cal N}$, when $|\epsilon|$ is far larger than
$10^2$ and thus $1/|\epsilon|\rightarrow 0$. For instance, we may
take $|\epsilon|\simeq 2.5\times 10^5 \simeq 10^3{\cal N}^{\,2}$
at the epoch of ${\cal N}\simeq 50$, and thus may have
${d\ln{|\epsilon|}\over d{\cal N}}={2\over {\cal N}}$, and thus
$n_s\simeq 0.96$, note that in this case in Eq.(\ref{ns}) ${2\over
|\epsilon|}\ll {d\ln{|\epsilon|}\over d{\cal N}}$ has been
neglected. Thus compared with that of $\epsilon$ being constant,
in which $|\epsilon|$ is required $\sim 10^2$ by the observations,
when $\epsilon$ is changed the range of $\epsilon$ constrained by
the observations may be more flexible, since in this case
$|\epsilon|$ may be quite large. We plot Fig.4, in which the
region of $|\epsilon|$ with respect to ${d\ln{|\epsilon|}\over
d{\cal N}}$, consistent with the observation, is given. We can see
that when $\epsilon$ is constant, i.e. ${d\ln{|\epsilon|}\over
d{\cal N}}=0$, in order to match the observation, $|\epsilon|$
must lie between about 20$\sim$100, which is a quite constrained
region. However, when ${d\ln{|\epsilon|}\over d{\cal N}}$
increases, $|\epsilon|$ may be taken as a larger value. From
Eq.(\ref{pxi}), for ${\cal P}_{(s\rightarrow \zeta)}\sim
10^{-10}$, we can see that a larger $|\epsilon|$ means a smaller
$h_e$, and thus a lower thermalization temperature, which is
actually good for an escaping from the monopole problem afflicting
the standard cosmology.
% is actually dependent of the rate that
%the \bd-branes are sent into the throat. When this rate is
%proximately constant, we will have $\epsilon$ constant, or
%$\epsilon$ will be time dependent. Thus in principle the spectrum
%may be blue or red, dependent of physical details of \bd-branes
%motion in the compactification bulk. To make the spectrum become
%slightly red, ${d\ln{|\epsilon |}\over d{\cal N}}$ is negative and
%larger than $2|\epsilon| $ is required, which means that
%$|\epsilon |$ must be increased rapidly with the decreasing of
%${\cal N}$. Thus it require that in per Hubble time the number
%$\Delta p$ of \bd-branes entering the throat should be more and
%more with the time.

\begin{figure}[t]
\begin{center}
\includegraphics[width=8cm]{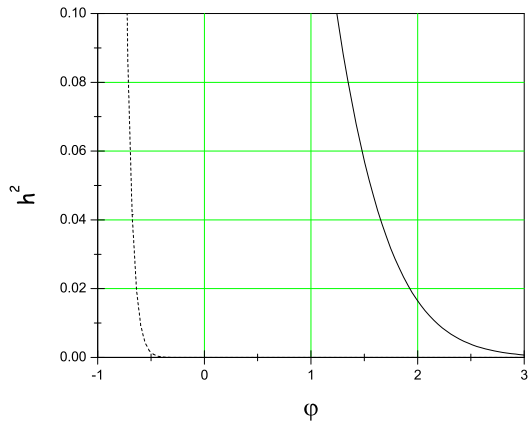}
\caption{The change of $h^2$ with respect to $\varphi$, and thus
${\cal N}$. The solid line is that with $|\epsilon|\simeq {\cal
N}$, and the dashed line is $|\epsilon|\simeq 10^3{\cal N}^{\,2}$.
}
\end{center}
\end{figure}

%We can see that if $|\epsilon|\sim 10^2$, the spectrum of entropy
%perturbation may be very naturally matched to recent observation
%\cite{WMAP}, since $n \equiv 1/|\epsilon |\sim 0.01$.

The potential $V(\varphi,s=0)$ of fields can be also obtained when
a relation between $\epsilon$ and ${\cal N}$ is given. We note
${\dot \varphi}=\sqrt{{\dot h}\over 4\pi}=\sqrt{|\epsilon|h^2\over
4\pi}$, and thus have \be {d{\cal N}\over d\varphi}\simeq
{\sqrt{|\epsilon|}\over 2\sqrt{\pi}}, \label{dndp}\ee where
$|\epsilon|\gg 1$ has been used, which means that by substituting
$\epsilon({\cal N})$, we will obtain a relation between $\varphi$
and $\cal N$. In addition, the function of $n({\cal N})$ can be
given by $-{1\over n} \simeq \epsilon - {d\epsilon\over d{\cal
N}}$. Thus after combining them, we can obtain a function
$n(\varphi)$, which then is submitted to Eq.(\ref{vphi1}) and thus
leads to the potential $V(\varphi,s=0)$. For a detail, taking
above example $|\epsilon|\simeq {\cal N}$, we will have ${\cal
N}\simeq {\varphi^2\over 16\pi}$ by Eq.(\ref{dndp}), and then
combining it with ${1\over n}\simeq {\cal N}$, where ${\cal N}\gg
1$ has been used, we can have ${1\over n}\simeq {\varphi^2\over
16\pi}$. Thus after substituting it into Eq.(\ref{vphi1}), we
obtain $V(\varphi)\sim \exp{(-{\varphi^2\over 2})}$. The similar
steps can be also applied to
another example $|\epsilon|\simeq 10^3{\cal N}^{\,2}$ mentioned%,
%for which we have $V(\varphi)\sim \exp{\left[-8\pi\times
%\exp{(\sqrt{10^3\over 4\pi}\varphi)}\right]}$
. By combining Eqs.(\ref{h}), (\ref{vphi1}) in which $s=0$ is
taken, and Friedmann equation, we can straightly obtain the change
of $h$,
%which is given by \be {h^2\over n^2}\equiv
%e^{-\int\sqrt{16\pi \over n}d\varphi},\label{hn}\ee
see Ref. \cite{Piao0705} for details.
%$h^2=n^2 e^{-\int\sqrt{16\pi \over n}d\varphi}$.
For both above examples, we plot Fig.5 for a comparison between
them, in which the calculations are exactly implemented without
any approximations. We can see that the case with larger
$|\epsilon|$ corresponds to a steeper change of local Hubble
parameter, which is an expected result.
%, whose figures for above
%examples is plotted in Fig.4. It is obvious that the more abrupt
%the change of $h$ is, the steeper the potential of field required
%to simulate its change is.

%\begin{figure}[t]
%\begin{center}
%\includegraphics[width=8cm]{nongaussii.eps}
%\caption{ The potential with respect to $\varphi$.}
%\end{center}
%\end{figure}

In addition, we also can obtain a slightly blue spectrum by
requiring ${d\ln{|\epsilon|}\over d{\cal N}}<0$ and
$\left|{d\ln{|\epsilon|}\over d{\cal N}}\right|>{2\over
|\epsilon|}$. This may be implemented e.g. by taking
$|\epsilon|\simeq {10^6\over {\cal N}}$ at the epoch of ${\cal
N}\simeq 50$, which leads to
%${2\over |\epsilon|} = 0.01$ and
%${d\ln{|\epsilon|}\over d{\cal N}} = \,-0.02$.
$n_s-1\simeq 0.01$ given by Eq.(\ref{ns}). Thus in principle we
could have any tilt required by the observations in such an island
universe thermalized in the landscape.

\subsection{The non-Gaussianity of curvature perturbation}

\begin{figure}[t]
\begin{center}
\includegraphics[width=8cm]{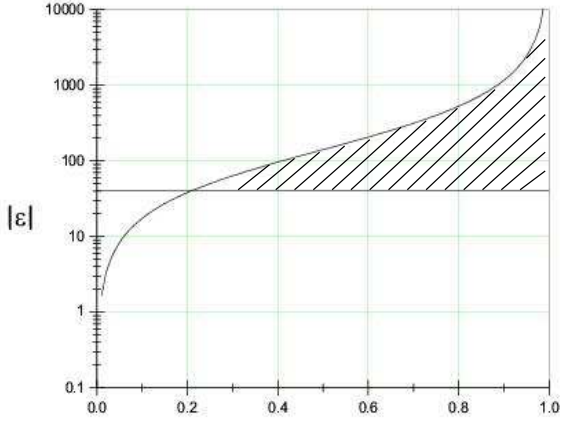}
\caption{ The region with tilt lines is that consistent with the
non Gaussiantity observation, in which the horizon axis is
$n_1/n_2$. $|\epsilon|$ is assumed as constant, and the upper
limit is determined by $f_{\rm NL} <100$ and the lower limit is
given by ${2\over |\epsilon|}\simeq 0.05$. }
\end{center}
\end{figure}

Then we will simply estimate the non-Gaussianity of this curvature
perturbation. Here the curvature perturbation is induced by the
entropy perturbation from $\delta s$ through Eq.(\ref{xi}), thus
in principle the non-Gaussiantity has two sources, one is the
cubic interaction terms of $s$ field, the other is the nonlinear
relation between $\delta s$ and $\zeta$, noting that Eq.(\ref{xi})
is only the result in linear approximation. Here we will estimate
the contribution to the non-Gaussianity from the cubic interaction
terms of $s$ field. To simplicity, we will assume that $n_1$ and
$n_2$ and thus $n$ are constant. In this case, the interaction
Hamiltonian is
%\be \left(\sqrt{n_1\over n_2}-\sqrt{n_2\over
%n_1}\right){\sqrt{16\pi}(\delta s)^3\over 3 \sqrt{n}(-t_e)^2} =
${\cal H}_{\rm int}=\alpha{(\delta s)^3\over \sqrt{n}(-t_e)^2}$,
%\label{alpha}\ee
where $\alpha =(\sqrt{n_1\over n_2}-\sqrt{n_2\over
n_1}){\sqrt{16\pi}\over 3}$ has been set, which can be obtained by
expanding the potential $V(\varphi, s)$ in Eq.(\ref{vphi1}) and
then taking the cubic part of $s$, in which Eqs.(\ref{h}) and
(\ref{vphi1}) and Friedmann equation have been used. Following
\cite{M}, and also recent \cite{CS}, the 3-point function of
$\delta s$ is given by \ba & &
<\delta s_{\vec{k}_1} \delta s_{\vec{k}_2} \delta s_{\vec{k}_3}> \nonumber\\
& = & -i\int_{-\infty}^{t_{\rm e}}<|[\delta s_{\vec{k}_1} \delta
s_{\vec{k}_2} \delta
s_{\vec{k}_3}, {\cal H}_{\rm int}(\lambda)]|>d\lambda +{\rm c.c.} \nonumber\\
& \cong &
(2\pi)^3{\delta\left(\sum_i\vec{k}\right)\sum_ik_i^3\over
\prod_ik_i^3}\cdot {\alpha\over \sqrt{2^3n}(-t_e)}{\cal
P}_{s}^{3/2}, \label{123}\ea which is calculated at the time of
thermalization, where the second line is obtained by only
reserving the leading order contribution for small $t_e$, and
$k_i$ is the amplitude of $\vec{k}_i$ and ${\cal P}_s\cong {1\over
2(-t_e)^2}$ is given by Eq.(\ref{ps}), since $a\eta_e\cong t_e$
for $\epsilon\ll -1$ and also $v\cong 3/2$.

Thus with Eq.(\ref{xi}), in super Hubble scale, the 3-point
function $<\zeta_{\vec{k}_1} \zeta_{\vec{k}_2} \zeta_{\vec{k}_3}>
$
%$\langle \zeta_{\vec{k}_1}(t_e)\zeta_{\vec{k}_2}(t_e)
%\zeta_{\vec{k}_3}(t_e)\rangle$
of curvature perturbation induced by $\delta s$ can be written as
\be \pm (2\pi)^3{\delta\left(\sum_i\vec{k}\right)\sum_ik_i^3\over
\prod_ik_i^3}\cdot {\alpha\over \sqrt{2^3n}(-t_e)}{\cal
P}_{(s\rightarrow \zeta)}^{3/2}, \label{gau1}\ee where $\pm$
appears since generally $\zeta\sim \pm \delta s$ and ${\cal
P}_{(s\rightarrow \zeta)}$ is given by Eq.(\ref{pxi}). The level
of non-Gaussianity is usually expressed in term of parameter
$f_{\rm NL}$ \cite{KS, BCZ} defined as
$\zeta(x)=\zeta_g(x)-{3\over 5}f_{\rm NL}\zeta^2_g(x)$. Thus
combining it with Eq.(\ref{gau1}), we have \ba f_{\rm NL}&\cong &
\mp {5\sqrt{2}\over 24}{\alpha\over {\sqrt n}(-t_e)} {1\over
\sqrt{{\cal P}_{(s\rightarrow \zeta)}}}\nonumber\\ & \cong & \mp
{5\sqrt{\pi}\over 24\sqrt{2}}\cdot { \alpha \over \Delta
\theta}|\epsilon|, \label{gau2}\ea where Eq.(\ref{pxi}), and also
$n\equiv {1\over |\epsilon|}$ when $n$ is constant, have been
used. Eq.(\ref{gau2}) shows that the non-Gaussianity is
proportional to $\epsilon|$, which is similar with that of usual
slow roll inflation model in which since $\epsilon\ll 1$ the
non-Gaussianity is generally very small. However, in island
cosmology, since $|\epsilon|\gg 1$, the non-Gaussianity is
generally quite large. This can also occur samely in ekpyrotic
comology \cite{CS, KMVW, BKO1}, in which $\epsilon \gg 1$ while
here $\epsilon\ll -1$, see also recent different study \cite{LS}.
However, note also that here $\alpha$ is actually dependent of
$n_2/n_1$. When $n_2=n_1$, we can obtain $\alpha=0$. Thus it is
also possible to make the level of non-Gaussianity be quite small
by adjusting $n_1$ and $n_2$, i.e. the ratio between both fields
contributing the background, which is actually determined by their
potentials. The observation gives $-36< f_{\rm NL} <100$
\cite{WMAP}. Thus an estimate of the upper limit of $\epsilon$ for
different $n_2/n_1$ may be obtained by noting $\Delta\theta\simeq
\arctan{\sqrt{n_2\over n_1}}$, which is plotted in Fig.6, in which
we take $f_{\rm NL} <100$. We can see that when
$n_2/n_1\rightarrow 1$, $\epsilon$ may be quite liberal, however,
in general case there is an upper limit for $\epsilon$. For
instance, taking $n_2/n_1\simeq 0.5$, we have $|\epsilon|\lesssim
150$. Thus dependent of different evolutions, the island universe
can have large or small non-Gaussianity, which makes it safely lie
within the observational bound. Here we do not consider the
contribution of the nonlinear relation between $\delta s$ and
$\zeta$ to the non-Gaussianty, however, this contribution is
actually the same order as that given by the cubic interaction
terms of $s$ field and thus hardly can affect our rough estimate
made here. We will back the detailed study of non-Gaussianty of
island universe in the future, noting the significant detection of
non Gaussianity \cite{YW}.

\section{Discussion}

\begin{figure}[t]
\begin{center}
\includegraphics[width=7cm]{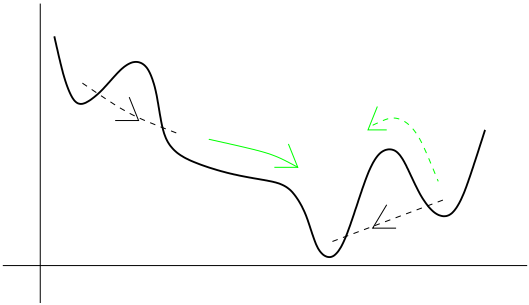}
\caption{In the eternal inflation leaded by the landscape, the
observable universes may be some of many thermalized regions
spawned within the eternally inflating background. They appear
either by a slow roll inflation after the nucleation of bubbles,
which is generally induced by the CDL instanton, like that given
by the black dashed line followed by the green solid line, or by a
straightly thermalization in new vacua without the slow roll
inflation, which may be induced by the HM instanton or others,
like that given by the green dashed line. In the meantime it can
also be expected that there are also lots of regions, which are
empty and not thermalized, like that given by the black dashed
line in right side.  }
\end{center}
\end{figure}

In this paper, we have illustrated and calculated the spectrum of
primordial perturbation of the island universe in the landscape.
The landscape can be depicted as the space of a set of fields. The
calculations implemented here closely capture this character of
landscape, since in this case the emergence of island will
inevitably involve the fluctuations of many fields, and thus the
entropy perturbation can be generated, which may induce the
curvature perturbation under certain condition. We showed a
detailed results of the spectrum of curvature perturbation induced
by the entropy perturbation and discussed that the parameter space
required by the observations should correspond to what change of
local Hubble parameter during the emergence of an island. We find
that in general case the results obtained can be well related to
those of slow roll inflation by a simple duality. In addition, we
also simply estimate the non-Gaussianity of perturbation. The
results shown here indicates that the island universe in the
landscape can be consistent with our real world.

Thus given the landscape, the observable universes may be some of
many thermalized regions spawned within the eternally inflating
background, which appear either by a slow roll inflation after the
nucleation of bubbles, followed by the reheating, or by a
straightly thermalization in new vacua without the slow roll
inflation, like islands, see Fig.7 for the illustration. Thus it
is interesting to ask how we know whether we live in an emergently
thermalized island or in a reheating region after slow roll
inflation inside bubble. The island universe generally has a large
non-Gaussianity, which can be distinguished by coming
observations, as has been shown here. However, a large
non-Gaussianity can be also achieved in some special inflation
models. In addition, it has been shown that in the island the
tensor amplitude is negligible on large scale \cite{Piao0706b,
Piao0506}, while there exists a large class of inflation models,
such as large field inflation model, with moderate amplitude of
tensor perturbation, see e.g. Ref. \cite{LR} for the various
inflation models. Thus it seems that the detection of a stochastic
tensor perturbation will be consistent with the inflation model,
while rule out the possibility that an straightly thermalized
region is regarded as our real world. However, low tensor
amplitude on large scale is also not conflicted with the inflation
model, e.g. some small field inflation models. Thus in this case
other distinguishabilities need to be considered. The bubble after
the nucleation described by the CDL instanton is generally
negatively curved, and thus the corresponding universe is an open
universe, while the island generated by the upward fluctuation may
be closed. Thus in this sense it seems that the curvature
measurement of our universe will be significant to make clear
where we live in. However, we still need to look for more distinct
signatures for island universe, which will be backed to in
intending works.

The emergence of island here is depicted as an upward fluctuation
with the null energy condition violation, which might be
interesting for the discussions of the initial conditions of
observable universe, see Refs. \cite{A02} and also recent
\cite{C07, I07}, since it might be closely relevant to the
solution of above issue. The emergence probability of island can
be approximately given by that of the HM instanton, which is
actually exponentially suppressed. However, in the bubble
nucleated by the CDL instanton, in order to have an universe like
ours, the slow roll inflation with enough period is generally
required.
%The observable universe would lie in a single nucleated bubble
%while the inflation continues forever outside of this bubble.
%However, the bubble universe nucleated is generally empty and
%negatively curved, which can hardly evolve to our real world. Thus
%to produce an universe like our, it seems that a period of slow
%roll inflation
%\cite{Guth, LAS}
%inside the bubble need to be applied \cite{FKMS}.
%, which reduces
%the curvature, provides the primordial density perturbation and
%large number of entropy required by the observable universe.
%However, to solve the horizon problems, the inflation is required
%to have an enough efolding number.
This only can be implemented by having a potential with a long
plain above its minimum, which obviously means a fine tuning,
since the regions with such potentials are generally expected to
be quite rare in a random landscape. While
%the island can become
%our observable universe is not dependent of whether there is the
%potential with a long plain.
the island of observable universe can actually emerge for any
potential, independent of whether the potential has a long plain,
as long as we can wait. Thus in principle the island of observable
universe can exist in any corner of landscape. This in some sense
brings us an interesting expectation that we might live in a
straightly thermalized ``island" in the landscape.

%However, the realistic evolution of universe with the landscape
%might show itself more fruitful than expected.

%The eternal inflation with the landscape makes some large and
%local quantum fluctuations with the null energy condition
%violation striding over the barriers between different vacua in
%the landscape and then straightly creating some thermalized
%regions in new vacua become possible. These thermalized regions or
%islands are filled with radiation and matter, which are similar to
%those after the reheating following a slow roll inflation, and
%thus will evolve with the conventional cosmology up to now.

%Thus it will be interesting to check whether there are some
%reliable and detailed predictions for island universe, compared
%with the observations.

%Thus it seems to be possible that we can live in ``Islands" in the
%landscape, but slightly not optimistic, since a full description
%for the phenomena with the null energy condition violation is
%still lacked for the moment, which is significant to complete the
%island universe model.

%The island universe is based on the fluctuation with the null
%energy condition violation, which is still a issue explored now.
%which might need some deep insights into the
%quantum gravity, we think that
%However, it can be expected that the results displayed here have
%phenomenally captured its some basic ingredients and possible
%predictions, which may be interesting and significant to the study
%of eternal inflation and landscape cosmology.

\textbf{Acknowledgments} The author thank Y.F. Cai for discussions
and careful comments. This work is based on the talk given in
KITPC workshop and is supported by KITPC, and also supported in
part by NNSFC under Grant No: 10775180, in part by the Scientific
Research Fund of GUCAS(NO.055101BM03), in part by CAS under Grant
No: KJCX3-SYW-N2.

\end{document}